\documentstyle[sprocl,epsf]{article}
\def\beq{\begin{equation}}
\def\brr{\begin{array}}
\def\err{\end{array}}
\def\eeq{\end{equation}}
\def\bea{\begin{eqnarray}}
\def\eea{\end{eqnarray}}

\begin{document}

\title{EXACT RENORMALIZATION GROUP APPROACH IN SCALAR AND FERMIONIC THEORIES}
\author{Yu. KUBYSHIN } 
\address{Institute of Nuclear Physics, Moscow State University \\
119899 Moscow, Russia}

\maketitle\abstracts{
The Polchinski version of the exact renormalization group equation 
is discussed and its applications in scalar and fermionic theories 
are reviewed. Relation between this approach and the standard 
renormalization group is studied, in particular the relation 
between the derivative expansion and the perturbation theory expansion is 
worked out in some detail.}

\section{Introduction}

As it is known (see an interesting review by Shirkov \cite{Sh1}) the basic 
idea of the renormalization group (RG) was formulated in an article 
by Stueckelberg and Petermann \cite{SP}. The existence of such 
group of transformations 
was related to the arbitrariness in the procedure of subtraction of 
ultraviolet divergencies in quantum electrodynamics. Functional equations 
for the propagators in the ultraviolet limit, corresponding to
the regularized Dyson transformations, were basically derived by Gell-Mann 
and Low \cite{GML}. Bogoliubov and Shirkov \cite{BSh1} 
unveiled the group nature
of these functional equations, established their relation to the
renormalization group by Stueckelberg and Petermann and derived
the functional group equations for propagators and vertices of 
quantum electrodynamics for a general case, i.e. massive case.
For example, in a renormalizable quantum field theory model
with one coupling constant $g$ and one mass $m$ (like 
$g \phi^{4}$ model in four dimensions) 
the functional equation for the invariant charge
$\bar{g}(x,y;g)$, with $x=p^{2}/\mu^{2}$ and $y=m^{2}/\mu^{2}$ being 
the ratios of the momentum and mass squared to 
the square of the normalization scale, has the following form: 
\beq 
\bar{g} (x,y;g) = \bar{g} \left(~\frac{x}{t}~, ~\frac{y}{t} ~, 
~\bar{g}(t,y;g) \right)~. 
\label{func} 
\eeq
This equation establishes the connection between the 
exact symmetry of Stueckelberg and Petermann and an approximate 
one of Gell-Mann and Low since in the massless limit $m=0$ it reduces 
formally to the result by Gell-Mann and Low \cite{GML}. Eq. (\ref{func}) 
reflects the exact symmetry of a solution which later became known 
as a functional self-similarity symmetry \cite{Sh82} (as it is 
relevant to the notion of self-similarity known in mathematical physics). 
From this equation the standard differential renormalization group 
equations in both Gell-Mann-Low and Callan-Symanzik form \cite{BSh,ChL} 
can be derived. We will call this approach 
just "renormalization group" (RG) in order to distinguish it from 
the approach originated from the works by Wilson and later by 
Polchinski, which is traditionally called "exact renormalization group" 
(ERG). 

The symmetry, underlying the RG, is 
the symmetry of a characteristic function $F(x; x_{0}, F_{0})$ of 
a physical problem ("solution of the problem") with respect to 
the change of the boundary condition 
$F(x_{0};x_{0},F_{0}) = F_{0}$ \cite{Sh1}. If we now consider 
$F(x;x_{1},F_{1})$, which is the {\it same} solution 
but with boundary condition settled at another point $x_{1}$, we arrive 
to the functional relation 
\[ 
F(x; x_{1}, F_{0}) = F(x; x_{1}, F(x_{1}; x_{0}, F_{0})).  
\]
Assuming that the function possesses the homogeneity property 
$F(x; x_{0}, F_{0} ) = f \left( \frac{x}{x_{0}}, F_{0} \right)$ with 
$f(1,F_{0}) = F_{0}$ we obtain for the function $f$ an equation of the 
type (\ref{func}). 

Another approach was developed in articles by Kadanoff and 
Wilson \cite{Ka,W}. It employes the idea of Wilson effective action, 
which is the action obtained by integrating out degrees of freedom 
with momenta $p^{2} \geq \Lambda^{2}$ in the defining 
functional integral. So, when the scale $\Lambda$ is reduced,  
the generating functional 
\beq
 Z_{\Lambda} (J) = \int \Pi_{|p| \leq \Lambda} 
    D\phi(p) \exp \left\{- S(\phi; \Lambda) -\int dp \ J \phi \right\}
\label{Z-def}
\eeq
includes integration over less and less modes, but the effective 
action $S (\phi; \Lambda)$ changes in such a way 
that the generating functional remains unchanged. In this sense, 
it describes the same physics for any $\Lambda$, 
that is the $S$-matrix elements and Green functions remain unchanged. 
These ideas were first developed for lattice statistical systems and were 
fruitfully used in condensed matter physics. Some re-formulations of this 
approach, which made possible its applications 
in quantum field theory, were 
made first by Wegner and Houghton \cite{WH} and Weinberg \cite{Weinberg} 
and later by 
Polchinski \cite{Po}. It is basically this approach which was 
called the exact renormalization group (ERG). In this article 
we discuss in detail the 
Polchinski version of the ERG (see a comprehensive review by Ball and 
Thorne \cite{BT} on this subject), though results obtained within other 
schemes will also be presented.  

The action can be characterized by a set $g = \{g_{2}, g_{4}, 
\ldots \}$ of coupling constants 
of all possible operators consistent with the symmetries of the 
system. The change of the scale $\Lambda  \rightarrow \Lambda / t$ 
combined with the corresponding change of the set $g \rightarrow 
R_{t} g$, where $R_{t}$ is some operator, can be viewed as a group 
transformation. Under such transformation the field changes as 
$\phi (p) \rightarrow \alpha_{t} \phi (tp)$. Then, for example, 
for the 2-point function, defined from (\ref{Z-def}) in a standard way,  
we have a relation of the type: 
\beq
   G_{2}\left(\frac{p}{\Lambda}, g(\Lambda) \right) = 
   \alpha_{t}^{2} G_{2} \left( \frac{tp}{\Lambda}, R_{t} g(\Lambda) \right).
\eeq
One can see similarity of this relation with the functional equation 
(\ref{func}). Having in mind this similarity, we can view the 
symmetry, underlying the Wilson renormalization group, as the invariance 
of the Wilson effective action with respect to the boundary condition
\beq
     S( \phi, \Lambda) |_{\Lambda = \Lambda_{0}} = S_{0}(\phi),
\eeq
where $S_{0}(\phi)$ is some "fundamental" action defined at the 
"fundamental" scale $\Lambda_{0}$. Because of 
this similarity one can expect that the RG and ERG approaches have much 
in common. This aspect was emphasized by Shirkov \cite{Sh1}. 

The first aim of this contribution is to give a short review of 
recent applications of the ERG in quantum field theory. Here we will 
limit ourselves to the case of scalar and fermionic theories only. 
Application of the ERG in gauge theories at the moment 
encounters difficulties (which are not unavoidable, in our opinion) 
due to the absence of a gauge invariant formulation \cite{gauge}. 
We will concentrate on results which are 
non-perturbative, though obtained within some approximation. 
This approximation will be the derivative expansion. 

The second aim is to work out some details on the relation between the 
RG and  the ERG. To much extent this was studied in the original 
paper by Polchinski \cite{Po} (for a detailed discussion see also the 
article by Ball and Thorne \cite{BT}). Explicit calculation of the 1-loop 
$\beta$-function and anomalous dimensions in scalar theory with the ERG 
method was done by Hughes and Liu \cite{HL}. In the present contribution 
we will examine the $\beta$-function calculated by the derivative 
expansion technique within the ERG and analyze its relation 
to the perturbative RG calculation. 

The article is organized as follows. In Sect.\ 2 we review the Polchinski  
version of the ERG equation in the case of scalar theory and describe the 
derivative expansion, which is a method used for non-perturbative 
calculations in this approach. In Sect.\ 3 we explain how the 
standard perturbative results can be obtained in the framework of  
the ERG approach and compare them with the results calculated within the 
derivative expansion. We will discuss the difference and  the 
relation between these two calculations. In Sect.\ 4 we review very 
briefly 
the main non-perturbative results obtained in scalar field theory 
within the ERG approach. Here calculations for various types of the ERG 
equations, not only for the Polchinski one, are summarized. In the 
last section the ERG equation for fermions and some first results obtained 
in the 2-dimensional Gross-Neveu type model are presented.

\section{The ERG equation}

Consider a $d$-dimensional scalar theory of the field $\Phi$ which 
is invariant under the $Z_{2}$-symmetry transformation 
$\Phi \rightarrow -\Phi$. Let us introduce the regularized 
generating functional \cite{Po,BT}
\beq 
    e^{-W(J;\Lambda)} = \int D\Phi e^{- S_{\Lambda}(\Phi;J)},
    \label{W-def}
\eeq 
where the action with the source term is given by
\bea
S_{\Lambda}(\Phi,J) & = & \int \frac{d^{d}p}{(2\pi)^{d}} \Phi(p) 
  P_{\Lambda}^{-1}(p) \Phi(-p) + S_{int}(\Phi; \Lambda)  \nonumber \\ 
       & + & \int d^{d}p J(p) Q_{\Lambda}^{-1}(p) \Phi(p) + 
         f_{\Lambda}.   \label{action}
\eea
The regularized propagator 
\beq
    P_{\Lambda}(p) = \frac{K\left( \frac{p^{2}}{\Lambda^{2}} \right)}
    {p^{2}}       \label{propag}
\eeq
is defined by introducing the regulating function $K(z)$ which is supposed 
to be decreasing fast enough when $z \rightarrow \infty$ and is 
normalized by $K(0) = 1$. The functions $Q_{\Lambda}(p)$ and 
$f_{\Lambda}$ 
are necessary for the consistency of the formulation and are  
to be determined. Note that the last term in Eq. (\ref{action}) does not 
depend on the field. 

In the functional integral above only those momentum modes which 
have $|p|$ lower or about $\Lambda$ are important, 
contributions of higher modes are suppressed by the 
the regulating function. Thus $\Lambda$ plays the role of a (smooth) 
upper cutoff. In this setting  
Wilson's idea is realized as follows: the effective 
action $S_{\Lambda}$ is such that while the scale $\Lambda$ varies the 
$S$-matrix elements or even the off-shell Green functions remain unchanged. 
This implies that the generating functional is a constant function of 
$\Lambda$: 
\beq
    \Lambda \frac{d W(J,\Lambda)}{d\Lambda} = 0.   \label{W-const}
\eeq  
The change of the propagator (the range of modes 
suppressed in the functional integral) with $\Lambda$ in the kinetic term 
is compensated by the change of the action of interaction and other terms 
in Eq. (\ref{action}) so that the whole functional integral, defining the 
generating functional, describes the same physics. Next step is to use 
the functional integral identity
\[
 \int D\Phi \frac{\delta}{\delta \Phi(p)} \left( \frac{1}{2} 
 \frac{\delta}{\delta \Phi(-p)} + P_{\Lambda}^{-1}(p) \Phi(p) + 
  Q_{\Lambda}^{-1}(p) J(p) \right) e^{-S_{\Lambda}(\Phi,J)} = 0.  
\]
Combining this identity with the condition (\ref{W-const}) one can derive 
the ERG equation for the effective action. Here we omit details of the 
derivation and present the result only. Before doing this  
let us introduce some notations. It is convenient to define the 
parameter 
\beq
       t = -\ln \frac{\Lambda}{\Lambda_{0}}, 
\eeq
where $\Lambda_{0}$ is some fixed scale. The scale of dimensionful 
objects is carried by $\Lambda$. So we can define the "dimensionless 
momentum" $q$ and the dimensionless field variable $\phi(q;t)$ as 
follows:
\beq
  q = \frac{p}{\Lambda},  \; \; \; \phi(q;t) = \Lambda^{1+ d/2} 
\Phi(p; \Lambda, \Lambda_{0}). 
\eeq
The dependence of the field on $t$ is characterized by the anomalous 
dimension $\eta$:
\beq
      \frac{\partial}{\partial t} \phi(q) = \frac{\eta}{2} \phi(q). 
\eeq
Then the part of the effective action (\ref{action}), which does not 
include the source term and the constant term $f_{\Lambda}$, takes the form
\beq
S[\phi] = \int \frac{d^{d}q}{(2\pi)^{d}} \phi(q) \frac{q^{2}}{K(q^{2})} 
\phi(-q) + S_{int}[\phi;t].      \label{action1}
\eeq
The ERG equation for it is 
\bea
\frac{\partial S}{\partial t} & = & - \int d^{d}q (2\pi)^{d} K'(q^{2}) 
  \left[ \frac{\delta^{2} S}{\delta\phi(q) \delta \phi(-q)} - 
    \frac{\delta S}{\delta \phi(q)} \frac{\delta S}{\delta \phi(-q)} 
\right] + S d   \label{ERG-eq}   \\
& + &  \int d^{d}q \phi(q) \frac{\delta S}{\delta \phi(q)} \left[
 \frac{\eta}{2} - 2q^{2} \frac{K'(q^{2})}{K(q^{2})} + 1 - \frac{d}{2} \right]
    - \int d^{d}q \phi(q) q_{\mu} \frac{\partial '}{\partial q_{\mu}} 
\frac{\delta S}{\delta \phi(q)}.  \nonumber
\eea
The prime in the last term means that the derivative 
does not act on the standard $\delta$-functions of the total 
energy-momentum 
conservation which appear in the action in the momentum representation. 
Only the first line is a non-trivial part, the rest of the terms 
just reflects the canonical dimensions of the objects of the action 
and the anomalous dimension of the field. 
For the condition (\ref{W-const}) to fulfill the function $f_{\Lambda}$
has to satisfy 
\[
\dot{f}_{\Lambda} = \int d^{d}p \left[ \tilde{Q}^{-2}(p^2) \frac{\dot{P}_{\Lambda}
 (p)}{P_{\Lambda}^{2}(p)} J(p)J(-p) - 
 \int d^{d}p \frac{\dot{P}_{\Lambda}(p)}{P_{\Lambda}(p)} \delta(0) \right]
\]
and $Q_{\Lambda}(p)=P_{\Lambda}(p) \tilde{Q}(p^2)$, where 
$\tilde{Q}(p^2)$ obeys the equation $\dot{\tilde{Q}}(p^2)=(\eta/2)\tilde{Q}
(p^2)$. Here the dot means differentiation with respect to $t$. 

The equation (\ref{ERG-eq}) is supposed to be supplied with the initial 
condition set at some scale $\Lambda_{0}$ or at $t=0$: 
\[
S_{int}[\phi;t]|_{t=0} = \tilde{S}_{int}[\phi] = 
\int d^{d}x \tilde{L}_{int}(\phi),
\]
where $\tilde{L}_{int}$ is essentially the bare Lagrangian. Then the ERG 
equation defines the running Lagrangian $L_{int}(\phi,t)$ and, 
correspondingly, the running action 
$S_{int}[\phi,t]=\int d^{d}x L_{int}(\phi,t)$, i.e. 
a trajectory in the space of all possible 
Lagrangians parametrized by $t$. Note that the equation (\ref{ERG-eq}) is 
exact and non-perturbative. One can observe certain similarity of it with 
a functional generalization of the RG equation to the case of 
Lagrangians of arbitrary type (including non-renormalizable 
Lagrangians) \cite{Kaz}.

The limit $t \rightarrow \infty$ describes either the situation 
with $\Lambda_{0}$ fixed and $\Lambda \rightarrow 0$, i.e. the 
limit of low characteristic energy, or the situation $\Lambda_{0} 
\rightarrow \infty$, i.e. the continuum limit of the model. In this way 
the ERG equation allows for non-perturbative studies of the 
continuum limit in quantum field theory.   

There are a few important issues that 
can be addressed by studying and solving this equation. First of all 
we can look for fixed point solutions $L_{int}^{*}(\phi)$ which 
some of the trajectories can approach as $t \rightarrow +\infty$ (here we 
mean fixed points for finite values of the coupling constants in the 
Lagrangian). The fixed points satisfy Eq. (\ref{ERG-eq}) with the zero 
l.h.s., $\dot{S}_{int}^{*}=0$, and this equation defines the value 
of the anomalous dimension $\eta=\eta^{*}$ for which such solution exists. 
The Gaussian fixed point $S_{int}^{*}=0$ gives an example 
of the trivial solution. 

Having found a fixed point solution we can study the theory in its vicinity. 
For this we represent the Lagrangian as an expansion in operators 
\beq
L_{int}(\phi,t) = L_{int}^{*}(\phi) + \sum_{n} {\cal O}_{n} e^{\lambda_{n}t}.
\eeq
The parameters $\lambda_{n}$ are called critical exponents and they are 
physical observables. 
The operators ${\cal O}_{n}(\phi)$, which correspond to 
$\lambda_{n}>0$, are called relevant operators and are important for the 
physics of the system in the vicinity of the fixed point when 
$t \rightarrow \infty$. The ERG equation (\ref{ERG-eq}) 
in principle allows to calculate the critical exponents and found 
the corresponding operators.   

Finally, one can try to solve 
the ERG equation for arbitrary $t$, i.e. find the complete 
renormalization group trajectory. 
There are also other interesting problems 
which can be considered in a non - perturbative way in the framework 
of this ap\-proach. These include 
bound states, Za\-mo\-lod\-chi\-kov $c$-function, etc.  

We would like to mention that in quantum field theory ERG equations of 
other types are also considered. Historically the first one was the 
Wegner-Houghton equation \cite{WH}. It was formulated for the sharp 
cutoff 
and was used in a number of articles for calculation of fixed points, 
critical exponents, flows, etc. in the scalar theory (see the article by 
Hasenfratz and Hasenfratz for one of the first detailed studies \cite{HH}). 
An approach based on an  
ERG equation with a sharp cutoff for the effective action $\Gamma_{eff}$ 
was developed by Morris \cite{Mo1}. Another version is the equation for the 
average effective action \cite{Wet}. Some of these results 
will be discussed in Sect. 4. 

\section{Approximations and relation between ERG and RG}

There are some special cases when the ERG equation (\ref{ERG-eq}) simplifies 
essentially and one can find its solutions. This happens, for example, 
in the theory with $N$-component scalar field in the limit 
$N \rightarrow \infty$ \cite{WH,N-large}.
In general it is not clear how the ERG equation (\ref{ERG-eq}) 
can be solved exactly. 
So we need to use an approximation to obtain solutions and to  
analyze them. For this it is useful to have an idea 
of a Lagrangian which captures essential features of the problem. 
In general, even if we start with a simple initial Lagrangian, like, for 
example, $\tilde{L}_{int}(\phi) = g\phi^{4}/4!$ at $t=0$, the running 
Lagrangian can include all possible operators constructed out of the 
field $\phi$ and its derivatives which are consistent with the 
symmetry of the problem. In the momentum representation the action can 
be written as an infinite series
\bea
S_{int}[\phi,t] & = & \frac{1}{2} \int \frac{d^{d}q}{(2\pi)^{d}}
   A_{2}(q,-q,t) \phi(q) \phi(-q) + \frac{1}{4!} \int \frac{d^{d}q_{1}
   d^{d}q_{2} d^{d}q_{3} d^{d}q_{4}}
  {(2\pi)^{3d}} \delta(\sum_{i=1}^{4} q_{i})  \nonumber \\
  & \times & 
  A_{4}(q_{1},q_{2},q_{3},q_{4},t) \phi(q_{1}) \phi(q_{2}) \phi(q_{3}) 
  \phi(q_{4}) + \ldots     \label{action2}
\eea
Then the ERG equation gives rise to an infinite system of coupled equations
\bea
\dot{A}_{2}(q,-q,t) & = & (2+\eta) A_{2}(q,-q,t) - 2q_{\mu}\frac{\partial}
 {\partial q_{\mu}} A_{2}(q,-q,t) + 2 K'(q^{2}) A_{2}(q,-q,t) \nonumber \\
 & - & \int\frac{d^{d}q_{1}}{(2\pi)^{d}} K'(q_{1}^{2})A_{4}(q,-q,q_{1},-q_{1},t),
  \nonumber \\
\dot{A}_{4}(q_{1},\ldots, q_{4},t) & = & (4-d+2\eta) A_{4} -
   \sum_{j=1}^{4} q_{j\mu}\frac{\partial}
 {\partial q_{j\mu}} A_{4} \nonumber \\ 
  & + & 2 \sum_{j=1}^{4} \left[ K'(q_{j}^{2}) A_{2}(q_{j},-q_{j},t)\right] 
 A_{4}(q_{1},\ldots q_{4},t)  \nonumber \\
  & - & \int\frac{d^{d}q}{(2\pi)^{d}} K'(q^{2})
  A_{6}(q_{1},\ldots , q_{4},q,-q,t),  \label{ERG-system} \\
 \ldots  & & \nonumber
\eea
Let us consider now the case $d=4$ as an example. 
The initial conditions, corresponding to one of the  
usual renormalization prescriptions in terms of the coefficient 
functions in (\ref{action2}), are settled at two scales \cite{HL,BT}: 

at $\Lambda = \Lambda_{0}$:
\bea
  A_{2}(q,-q,\Lambda_{0})& =& \rho_{1}(\Lambda_{0}) + 
  q^{2}\rho_{2}(\Lambda_{0});
   \nonumber  \\
 A_{4}(q_{1}, \ldots , q_{4},\Lambda_{0}) & = & g(\Lambda_{0}) \equiv g_{B}; 
   \nonumber \\
  A_{2j}(q_{1}, \ldots , q_{2j},\Lambda_{0}) & = & 0 \; \; \; 
   \mbox{for $j \geq 3$},   \label{renorm1}
\eea
that gives the standard form for the bare Lagrangian where, of course, 
$g_{B}$ is the bare coupling constant; 

and at some physically relevant scale (renormalization point) 
$\Lambda = \mu_{R}$:
\bea
A_{4}(0,0,0,0,\mu_{R}) & = & g(\mu_{R});  \nonumber \\
\rho_{1}(\mu_{R}) = 0, & & \; \; \; \rho_{2}(\mu_{R})=0.  \label{renorm2}
\eea  
In the formulas above for illustrative purposes we indicated the  
dependence on the scale as the dependence on $\Lambda$ and not on $t$ as 
before.
  
One of the known ways to solve the ERG equation is to use the perturbative 
expansion. We assume that all coefficient functions can be 
represented as power series in $g_{R} \equiv g(\mu_{R})$. After 
a bit lengthy but straightforward calculation one can get
solutions for $A_{2j}$. In particular, 
\bea
g(\Lambda;\mu_{R},g_{R}) & \equiv & A_{4}(0,0,0,0,\Lambda; \mu_{R}, g_{R}) 
        \nonumber \\
  & = & g_{R} - 3 g_{R}^{2} \int \frac{d^{4}p}{(2\pi)^4} \frac{1}{p^2}
  \frac{1}{p^2} \int_{\mu_{R}}^{\Lambda} d\Lambda' 
  \left(\frac{d}{d \Lambda'} 
  K \left(\frac{p^{2}}{\Lambda'^{2}} \right) \right) 
  \left[ K \left(\frac{p^{2}}{\Lambda'^{2}} \right) - 
  K \left(\frac{p^{2}}{\Lambda_{0}^{2}} \right) \right] \label{g-PT0}  \\
   & = & g_{R} - 3 g_{R}^{2} \int \frac{d^{4}p}{(2\pi)^4} \frac{1}{p^2} 
 \frac{1}{p^2} \left\{ \frac{1}{2} \left( K^{2}\left(\frac{p^{2}}{\Lambda^2} 
\right) -  K^{2}\left( \frac{p^{2}}{\mu_{R}^{2}}\right) \right) \right.
              \nonumber \\
 & - &  \left.  \left( K \left(\frac{p^{2}}{\Lambda^2} 
\right) - K\left( \frac{p^{2}}{\mu_{R}^{2}}\right) \right) K \left( \frac{p^2}
{\Lambda_{0}^{2}} \right) \right\} + {\cal O}(g_{R}^{3}).
 \label{g-PT}   
\eea
In this formula we indicated the dependence on the initial condition 
(i.e. the renormalization point) explicitly. Taking the limit 
$\Lambda_{0} \rightarrow \infty$ we get the standard $\beta$-function 
in the 1-loop approximation \cite{HL}: 
\beq
\beta(g_{R}) \equiv \Lambda \frac{d}{d \Lambda} g(\Lambda; \mu_{R}, g_{R})
= \frac{3}{16 \pi^2} g_{R}^{2} + {\cal O}(g_{R}^{2}).  \label{beta}
\eeq
 
Can we get non-perturbative solutions of the ERG equation? At the moment 
the available systematic techniques for this are the derivative expansion and 
its modifications. The idea is to represent the action (here in 
the coordinate representation) as 
\bea
S_{int}[\phi,t] & = & \int d^{d}x \left[ V(\phi(x),t) + 
  (\partial_{\mu} \phi)^2 U(\phi(x),t) \right.    \nonumber \\
 & + & \left. (\partial_{\mu} \phi)^4 H_{1}(\phi (x),t) + 
 (\partial_{\mu}^2 \phi)^2 H_{2}(\phi(x),t) + \ldots \right], \label{der-exp}
\eea
where $V$, $U$, $H_{1}$, $H_{2}$, $\ldots$ depend on the field but not on 
its derivatives. In the momentum representation Eq. (\ref{der-exp}) 
corresponds to the expansion in powers of momenta, so one can hope that 
using a few first terms of this expansion is justified if we 
consider effects at low momenta (there may be some complications in 
the case of the sharp cutoff \cite{Mo2}). The function $V(\phi,t)$ is the 
effective (local) potential of the theory. To solve the ERG equation 
(\ref{ERG-eq}) approximately we truncate the derivative expansion 
(\ref{der-exp}) and substitute the truncated action into the ERG 
equation thus getting equations for the coefficient functions. 
For example, if we consider just the leading and next-to-leading 
terms we have \cite{BHLM}: 
\bea
\dot{V} & = & -\alpha V'' - 2\beta U'' + \gamma (V')^2 + d \cdot V 
 \left(1 - \frac{\eta}{2} - \frac{d}{2} \right) + \phi V'    
     \label{ERG-d1} \\
 \dot{U} & = & - \alpha U'' + \delta (V'')^{2} + 4 \gamma U V'' + 
  2 \gamma U' V' - \eta U  + \left(1 - \frac{\eta}{2} - \frac{d}{2}\right)
  \phi U' - \frac{\eta}{2},    \label{ERG-d2}
\eea
where the prime means differentiation with respect to the field $\phi$. 
The equation depends explicitly on the parameters characterizing the 
regulating function:
\beq
\alpha = \int  \frac{d^{d}q}{(2\pi)^d} K'(q^2), \; \; \; 
\beta = \int  q^2 \frac{d^{d}q}{(2\pi)^d} K'(q^2), \; \; \;
 \gamma = K'(0), \; \; \; \delta = K''(0). \label{scheme}
\eeq
This is similar to the dependence on the renormalization scheme in the 
RG. However, there is a problem of the breaking of the reparametrization 
invariance by the derivative expansion that gives rise additional 
dependence on the regulating function \cite{repinv,Mo8}.    

One way is to solve the system (\ref{ERG-d1}), (\ref{ERG-d2}) 
numerically. Similar systems for the ERG equations 
of other types were also studied \cite{Mo3}. These results will be discussed 
in the next section. 

Another way is to do further approximation, namely, to expand the 
functions $V$, $U$, etc. in powers of fields:
\bea
 V(\phi,t) & = & a_{1}(t) \phi^{2} + a_{2}(t) \phi^4 + a_{3}(t) 
   \phi^6 + \ldots,      \nonumber \\
 U(\phi,t) & = &  b_{2}(t) \phi^4 + b_{3}(t) \phi^6 + \ldots  
  \label{VU-polynom}
\eea
Then the system (\ref{ERG-d1}), (\ref{ERG-d2}) becomes  
the following set of flow equations \cite{CKM1}
\bea
\dot{a}_{1} & = & (2+\eta)a_{1} - 12 a_{2} - 6b_{2}/s_{1} + 4a_{1}^{2}, 
  \nonumber \\
\dot{a}_{2} & = & (4-d+2\eta)a_{2} - 30 a_{3} - 10b_{3}/s_{1} + 16a_{1}^{2}, 
  \nonumber \\
\dot{a}_{3} & = & (6-2d+3\eta)a_{3} + 24 a_{1} a_{3} + 16a_{2}^{2},
  \nonumber \\
\dot{b}_{2} & = & (2-d+2\eta)b_{2} + 16 a_{1}a_{2} + 16 a_{1} b_{2}
 - 20b_{3}, 
  \nonumber \\
\dot{b}_{3} & = & (4-2d+3\eta)b_{3} + \frac{192}{5} a_{2}b_{2} + 
  24 a_{1}b_{3} +  24 a_{1} a_{3} + \frac{144}{5} a_{2}^{2} 
  \label{ERG-p1}
\eea
with the relation 
\beq
   0 = \eta - s_{2} (12 b_{2} - 8 a_{1}^2).   \label{eta-eq}
\eeq
Here we introduced the combinations of the scheme parameters 
$s_{1} = \alpha \gamma / \beta \delta$ and $s_{2}=\delta / \gamma^2$.
The last equation arises because the normalization of 
the kinetic term is fixed. Let us mention that 
Eqs. (\ref{ERG-p1}) are basically $\beta$-functions for the  
coupling constants of corresponding operators.

The leading order of the derivative expansion with subsequent polynomial 
approximation of the potential $V(\phi,t)$ in scalar theories for the 
Wegner-Houghton ERG equation was studied in detail \cite{MOP,HKLM,Mo2}. 
We summarize some of the results in the next section.
Now we are going to illustrate the 
relation between the solutions within the derivative expansion in the ERG 
and the perturbative results in the RG (for $d=4$). For this purpose let us 
solve Eqs. (\ref{ERG-p1}), (\ref{eta-eq}) perturbatively, i.e. presenting 
the coefficients $a_{i}(t)$ and $b_{i}(t)$ and $\eta$ as series 
in powers of $g(t) \equiv 4! a_{2}(t)$, the coupling constant 
of the $\phi^4$-interaction. After a simple calculation up 
to the next-to-leading order ${\cal O}(\partial ^{2})$ of the 
derivative expansion one gets 
\beq
\beta(g) = - \frac{d}{dt} g(t) = 6(\alpha \gamma + \frac{1}{2} \beta \delta + 
  \ldots) g^2 + {\cal O}(g^3).       \label{beta-d1}
\eeq
We see that the expression (\ref{beta}) for the $\beta$-function to the 
order $g^2$ is not recovered within this approximation. Moreover, the 
result (\ref{beta-d1}) for the first coefficient of the $\beta$-function 
depends on the scheme. Of course, what happens is that there are   
contributions with higher derivatives (or with higher powers of the momenta) 
in the expansion (\ref{der-exp}) which give rise to further terms 
in the equations (\ref{ERG-p1}), (\ref{eta-eq}). 
In the perturbative solution these terms provide further contributions 
to the expression (\ref{beta-d1}), also to the $g^2$ coefficient (the 
dots in that expression stand for these contributions). To see the form 
of these contributions it is useful to re-write Eq. 
(\ref{beta-d1}), using the definitions (\ref{scheme}), as follows 
\[
\beta(g) = 6 g^{2} \int \frac{d^{4}q}{(2 \pi)^{4}} K'(q^2)
  \left[ K'(0) + \frac{1}{2} q^2 K''(0) + \ldots \right] + {\cal O}(g^3).
\]
Analyzing the relevant contributions of higher derivatives one can 
show that when they are taken into account the expression above sums up to 
\beq
\beta(g) = 6 g^{2} \int \frac{d^{4}q}{(2 \pi)^{4}} K'(q^2) \frac{1}{q^2}
  \left[ K(q^2) - 1 \right] + {\cal O}(g^3).
  \label{beta-d2}
\eeq
The integral can be calculated for an arbitrary regulating function 
and one obtains 
\[ 
   \beta(g) = \frac{3}{16 \pi^2} g^2 + {\cal O}(g^3),
\]
the standard result (\ref{beta}). In this calculation we used only that 
$K(\infty)=0$ and the normalization $K(0)=1$. 

This example illustrates the relation between the derivative 
expansion of the ERG and the perturbation theory of the RG. 
On one hand the derivative 
expansion contains only a part of the contribution of  
the weak coupling perturbative expansion to a given order in $g$. 
It can be obtained by making the Taylor expansion of a part of 
terms in the integral over internal momenta of the corresponding 
Feynman diagram. Indeed, consider expression (\ref{g-PT0}) or 
(\ref{g-PT}), which can be 
interpreted as a contribution of the 1-loop diagram with zero external 
momenta regularized according to Eq.\ (\ref{propag}). 
If we take the limit $\Lambda_{0} \rightarrow \infty$ 
and differentiate (\ref{g-PT0}) with respect to $\Lambda$, according 
to the definition (\ref{beta}), we obtain precisely the expression 
(\ref{beta-d2}). Now, if for infinite $\Lambda_{0} $ we expand  
the part in the square brackets in powers of $p^2$ (the 
integral still remains finite), we get the expression which 
after differentiation with respect to $\Lambda$ provides the result 
(\ref{beta-d1}) of the derivative expansion in the next-to-leading order. 

On the other hand the derivative 
expansion, when is not limited to the weak coupling expansion, 
contains non-perturbative information, that makes it and the 
ERG equation in general to be rather valuable tools. 
We will discuss the results of 
non-perturbative calculations in the next sections. 

\section{Scalar theory: main results}

Here we discuss some of the main results of the studies within  
the ERG approach in the scalar theory 
of one field with $Z_{2}$-symmetry (symmetry under the transformation 
$\Phi \rightarrow - \Phi$) on the $d$-dimensional Euclidean space.  
We collect results obtained by various authors and for different 
versions of the ERG equation. 
On general grounds different versions should give physically equivalent 
results, and concrete calculations confirm this (see the discussion of 
fixed point solutions and critical exponents below). However, no 
rigorous proof of the equivalence of different approaches has been 
given so far. 

First, there were numerous studies of fixed point solutions for various 
$d$. We discuss them in turn. 

1. \underline{ Fixed points: $d=4$.} Hasenfratz and Hasenftatz 
performed a study of the Wegner-Houghton version of 
the ERG equation (analog of Eq.\ (\ref{ERG-eq})) 
numerically in the leading approximation ${\cal O}(\partial ^{0})$of the 
derivative expansion (local potential approximation) and showed that 
it has no non-trivial fixed point solutions \cite{HH}.

2. \underline{ Fixed points: $d=3$.} In the same article it was 
showed, also numerically, that in order ${\cal O}(\partial ^{0})$  
the Wegner-Houghton equation has one non-trivial fixed point solution, 
the known Wilson-Fisher fixed point \cite{HH}, which 
is in the universality class of the three-dimensional Ising model. 
This fixed point was also studied in the ${\cal O}(\partial ^{2})$ 
approximation, both  
for the Polchinski type ERG equation (\ref{ERG-eq}) \cite{BHLM} 
and for the Wegner-Houghton equation \cite{Mo8,Mo3} . 

As an example, let us review the calculation for the Polchinski type 
ERG equation. 
To the order ${\cal O}(\partial ^{0})$ of the derivative expansion the fixed 
point equation is Eq. (\ref{ERG-d1}) without  
the term $2\beta U''$ and the l.h.s. $\dot{V} = 0$.  
It was solved numerically with two boundary conditions at $\phi = 0$:
\[
 V'(0) = 0, \; \; \; V''(0) = \rho, 
\]
the first one just reflecting the symmetry of the theory. 
It was shown that a fixed point solution $V^{*}(\phi)$ regular for all 
finite values of $\phi$ exists only for a special value of $\rho = 
\rho^{*} = -0.2286 \ldots $ In this approximation the fixed point solution  
$V^{*}(\phi)$ does not depend on the scheme parameters and the value of 
the anomalous dimension at the fixed point is $\eta^{*} = 0$. In 
the next-to-leading approximation ${\cal O}(\partial ^{2})$ the 
system of fixed point equations , 
i.e. Eqs. (\ref{ERG-d1}), (\ref{ERG-d2}) with zero left hand sides, 
was solved numerically with the following boundary conditions for $V(\phi)$ 
and $U(\phi)$ at $\phi = 0$:
\[
V'(0) = 0, \; \; \; V''(0) = \rho, \; \; \; 
 U(0) = 0, \; \; \; U''(0) = 0.
\]
A regular non-trivial solution exists and is unique 
only for a special value of $\rho = \rho^{*}$ and the anomalous dimension 
$\eta = \eta^{*} \neq 0$, but now $\rho^{*}$ and $\eta^{*}$ 
depend on the scheme parameters (\ref{scheme}). 
 
3. \underline{ Fixed points: $2 < d \leq 4$.} 
For this range of dimensions the Wegner-Houghton fixed point equation was 
analyzed in the leading order of the derivative expansion with 
subsequent approximation of the local potential $V^{*}(\phi)$ by a 
polynomial in powers of the field \cite{HKLM,HKLM1}: 
\beq 
  V_{M}^{*} (\phi) = a_{1}^{*} \phi^{2} + a_{2}^{*}\phi^{4} + \ldots 
  + c_{M}^{*} \phi^{2M}. \label{V-poly}
\eeq
In this case the fixed point ERG equation reduces to a system of 
$M$ algebraic equations for the coefficients $a_{i}^{*}$. As before, this 
approximation gives the anomalous dimension $\eta^{*}=0$, nevertheless 
it captures some general features of the multicritical fixed points 
below $d=4$. In particular, for approximation (\ref{V-poly}) the 
system shows the existence of the first $(M-1)$ upper critical 
dimensions $d_{k} = 2k/(k-1) = 4, 3, 8/3, \ldots$ for $k=2,3,4, \ldots, M$. 
Then, at $d=d_{k}$ the trivial Gaussian fixed point 
$V_{M}^{*} = 0$ is a branching point of a new non-trivial fixed point and 
the Gaussian one below $d_{k}$. 

A general feature of the polynomial approximation is the appearance 
of numerous spurious solutions. For example, for $d=3$ from the 
numerical results, discussed above, we know that there is only one 
non-trivial fixed point, the Wilson-Fisher fixed point. But if we 
approximate the potential $V^{*}(\phi)$ by a polynomial 
$V_{M}^{*}(\phi)$ the system can have till $M$ real valued non-trivial 
solutions for the coefficients $a_{i}^{*}$, all of them but one being 
spurious. The problem of spurious solutions can be solved by 
analyzing solutions of subsequent approximations with various 
$M$ and selecting the stable ones, which represent the true fixed points. 

The polynomial approximation permits an analytical study and reproduces 
some of the qualitative features of the structure of the fixed points. 
However, arguments were presented which show 
that the polynomial approximation is not convergent \cite{Mo6}, 
i.e. for a given $\phi$ $|V_{M}^{*} (\phi) - V^{*}(\phi) |$ 
approaches small but non-zero value 
as $M \rightarrow \infty$. 

4. \underline{ Fixed points: $d=2$.} 
The case of $d=2$ dimensions was studied by Morris \cite{Mo8,Mo7}. 
At ${\cal O}(\partial^{2})$ order of the derivative expansion he found 
first 10 multicritical fixed points of an infinite series which corresponds 
to unitary minimal models of conformal field theory and whose existence was 
conjectured by Zamolodchikov.  

5. \underline{Critical exponents}. 

To calculate the critical exponents at a given fixed point we 
expand the potentials $V$, $U$, etc. around the fixed point solution,
\beq
  V(\phi,t) = V^{*}(\phi) + \delta V (\phi, t), \; \; \; 
  U(\phi,t) = U^{*}(\phi) + \delta U (\phi, t), \ldots, 
\eeq 
linearize the ERG equations (\ref{ERG-d1}), (\ref{ERG-d2}) with respect 
to the linear deviations $\delta V$, $\delta U$, etc. and represent these 
deviations in the vicinity of the fixed point as
\beq
   \delta V (\phi,t) = \sum_{n} v_{n}(\phi) e^{\lambda_{n} t}.
\eeq
The critical exponents $\lambda_{n}$ can be found as eigenvalues of the 
linearized system of the ERG equations. 

Most of the study was done for critical exponents of the 
Wilson-Fisher fixed point for $d=3$. There is one positive critical 
exponent $\lambda_{1}$ and, correspondingly, one relevant operator, 
the rest of $\lambda_{n} < 0$ $(n \geq 2)$. We give a summary of the results 
of various calculations of $\nu \equiv 1/\lambda_{1}$, characterizing the 
critical exponent of the relevant operator, $w = -\lambda_{2}$ for 
the first irrelevant operator, and the anomalous dimension 
$\eta$ in Table 1.  
\begin{table}[ht]
\renewcommand{\arraystretch}{1.5}
\hspace*{\fill}
\begin{tabular}{|l|l|l|l|} \hline
Approach and approximation & $\eta$ & $\nu$ & $w$ \\  \hline
Wegner-Houghton eq. \cite{HH}, ${\cal O}(\partial^{0})$, numerically 
& 0 & 0.687 & 0.595   \\ \hline
Wegner-Houghton eq. \cite{HKLM}, ${\cal O}(\partial^{0})$, polynom., $M=7$ & 
0 & 0.657 & 0.705  \\  \hline
Eq. for the Legendre effective action \cite{Mo2} & & & \\  
${\cal O}(\partial^{0})$, numerically  & 0     & 0.660 & 0.628 \\ 
${\cal O}(\partial^{2})$, numerically  & 0.054 & 0.618 & 0.897 \\  \hline
Polchinski eq.\cite{BHLM} & & & \\  
${\cal O}(\partial^{0})$, numerically  & 0     & 0.649 & 0.66 \\ 
${\cal O}(\partial^{2})$, numerically  
& 0.019-0.056 & 0.616 - 0.637 & 0.70 - 0.85 \\ \hline
World best estimates  & 0.035(3) & 0.631 (2) & 
 0.80(4)  \\  \hline
\end{tabular}
\hspace*{\fill}
\renewcommand{\arraystretch}{1}
{\caption[Results of calculations]
{Results of calculations of the anomalous dimension $\eta$ and the 
critical exponents $\nu$ and $w$ for the Wilson-Fisher fixed point 
at $d=3$ by various authors for different versions of the ERG 
equation. The entries of the last row (taken from the article by 
Morris \cite{Mo2,Mo8}) were obtained by averaging the world best 
estimates \cite{World}.}}
\label{t1}
\end{table}
One can see that the results of different schemes and approximations 
are in a reasonable agreement. In cases when the 
characteristic under consideration is scheme dependent the intervals 
of values, which correspond to certain ranges of values of the 
scheme parameters, are indicated (the reader is referred to the original 
articles for details).

6. \underline{ Exact flow}.

Critical exponents characterize the flow very close to a particular 
fixed point. Within the approximation, considered here, one can also 
study the flow globally. 
Thus, the Wegner-Houghton flow equation was studied in the 
leading order of the derivative expansion with the potential $V(\phi,t)$ 
being approximated by the polynomial  
\beq 
  V_{M} (\phi,t) = \frac{1}{2} c_{1}(t) \phi^{2} + \frac{1}{4} 
   c_{2}(t)\phi^{4} + \ldots + \frac{1}{2M} c_{M}(t) \phi^{2M} 
  \label{V-poly1}
\eeq
(cf. (\ref{V-poly})) \cite{HKLM}. We refer 
the reader to this article for details.   

7. \underline{ Zamolodchikov $c$-function}.

Another interesting application of the ERG equation is the 
calculation of the approximate Zamolodchikov $c$-function 
which characterizes the geometry of the space of interactions of a 
given system.

The Zamolodchikov $c$-function $C(t)$ is a function which decreases 
monotonically along the flow (i.e. as the flow parameter $t$ grows) 
and which is 
stationary at fixed points of the theory: 
\[
   \frac{d C (t)}{d t} |_{\mbox{fixed point}} = 0.
\]
Zamolodchikov proved the existence of such function, which is unique 
up to a multiplicative factor, in two-dimensional 
unitary theories \cite{Zam}. There were some attempts to prove the 
existence of such function or construct it perturbatively in other 
cases \cite{c-function}. Here we will 
show that for the Wegner-Houghton 
ERG equation the flow for the potential, approximated by the 
polynomial (\ref{V-poly1}), is gradient and permits a $c$-function 
description. The former means that the beta-functions of the couplings 
$c_{i}(t)$, parametrizing the potential, are gradients of some 
function $C(c_{1},c_{2},\ldots)$:
\beq
  \frac{d}{dt} c_{i}(t) = - g_{ij} \frac{\partial C}{\partial c_{j}}, 
     \label{grad}
\eeq
where $g_{ij}$ is a positive definite metric in the space of coupling 
constants.  This implies that the set of renormalization flows 
is irreversible \cite{WZ}. Then $C$ is the $c$-function: 
\[ 
    \frac{d C}{dt} = \frac{\partial C}{\partial c_{i}} \frac{ dc_{i}}{dt} 
    \leq 0.
\]
The potential $V(\phi,t)$ in the polynomial approximation 
(\ref{V-poly1}) with $M=2$ for the Wegner-Houghton ERG equation 
was analyzed and the flow was shown to be gradient in this 
approximation \cite{HKLM} . 
In this case the system of flow equations consists of two equations: 
\bea
 \dot{c}_{1}(t) & = & 2 c_{1} + \frac{6c_{2}}{1+c_{1}},  \nonumber \\
  \dot{c}_{2}(t) & = & (4 - d) c_{2} - \frac{18 c_{2}^{2}}{(1+c_{1})^{2}} 
   \nonumber  
\eea
and can be cast into the form (\ref{grad}) with 
\begin{displaymath}
  g_{ij} = \frac{1}{(1+c_{1})^{2}} \left( \begin{array}{cc}
1 & 0 \\ 0 & \frac{(4-d)}{3} c_{2} 
\end{array} \right) 
\end{displaymath}
and 
\beq
 C(c_{1},c_{2}) = - \frac{1}{2} (1+c_{1})^{4}  + \frac{2}{3} (1+c_{1})^{3}  
 - 3 c_{2}(1+c_{1})^{2} + \frac{27 c_{2}^{2}}{4-d}. \label{c-func}
\eeq
The $c$-function (\ref{c-func}) is stationary at the 
fixed points. For example, for $d=3$ it has the maximum at the Gaussian 
fixed point $(c_{1}=0,c_{2}=0)$ with $C(0,0)=1/6$ and a saddle point 
at the Wilson-Fisher fixed point $(c^{*}_{1}=-1/7, c_{2}^{*}=2/49)$ 
with $C(c_{1}^{*},c_{2}^{*})=252/2401 \approx 0.105$. 

\section{Fermionic theory: first attempts.}

In this section we discuss briefly the results of ERG studies carried 
out for a two-dimensional fermionic model \cite{CKM,Co}. 

The Polchinski version of the ERG equation for pure fermionic theories can 
be derived in the same way as the one for the scalar theories (with another 
functional identity, of course). We start with a general action given by 
\beq
  S = \int d^{d}p \bar{\Psi}(p) P_{\Lambda}^{-1} (p) \Psi (-p) + 
  S_{int}(\Psi,\bar{\Psi},\Lambda), \label{action-ferm}
\eeq
with the regularized propagator equal to
\[
    P_{\Lambda} (p)= i \hat{p} \frac{K\left(\frac{p^{2}}{\Lambda^{2}} 
    \right)}{p^{2}}.
\]
The analog of Eq. (\ref{ERG-eq}) is:
\bea
  \frac{\partial S}{\partial t} & = & 2 \int d^{d}q (2\pi)^{d} 
   K'(q^{2}) \left[ \frac{\delta S}{\delta \psi (q)} i\hat{q} 
   \frac{\delta S}{\delta \bar{\psi} (-q)} - 
   \frac{\delta }{\delta \psi (q)} i\hat{q} 
   \frac{\delta S}{\delta \bar{\psi} (-q)} \right]  \nonumber \\
   & + & d \cdot S + \int d^{d}q \left( \frac{1-d+\eta}{2} - 2q^{2} 
   \frac{K'(q^{2}}{K(q^{2})} \right) \left(
   \bar{\psi} (q) \frac{\delta S}{\delta \bar{\psi} (q)} + \psi (q) 
   \frac{\delta S}{\delta \psi (q)} \right) \nonumber \\
   & - & \int d^{d}q \left( \bar{\psi}(q) q^{\mu} \frac{\partial '}
   {\partial q^{\mu}} \frac{\delta S}{\delta \bar{\psi} (q)} + 
    \psi(q) q^{\mu} \frac{\partial '}
   {\partial q^{\mu}} \frac{\delta S}{\delta \psi (q)}  \right).
    \label{ERG-ferm}
\eea
Similar to the scalar case, $\eta$ is the anomalous dimension, $\psi$ and 
$\bar{\psi}$ are dimensionless fermionic fields, $q$ is the dimensionless 
momentum and $t$ is the renormalization flow parameter $t = - \ln (\Lambda / 
\Lambda_{0})$. 

A concrete theory, studied within this approach, is the two-dimensional 
Euclidean chiral Gross-Neveau type model. This is a 
theory of fermions with $N$ flavours described by 
2-component spinors $\psi^{a}(q)$, $\bar{\psi}^{a}(q)$, $a=1,2, 
\ldots, N$, with $SU(N)$ symmetry with respect to flavour indices of the left 
and right fields. The $\gamma$-matrices are given by Pauli matrices 
$\gamma_{1} = \tau_{1}$, $\gamma_{2} = \tau_{2}$, and $\tau_{3}$ plays 
the role of the chiral matrix $\gamma_{5}$.
It can be shown that the most general action 
respecting such symmetries can be constructed out of 
the following operators:
\beq
  S(q_{1},q_{2}) = \bar{\psi}^{a}(q_{1}) \psi^{a}(q_{2}), 
  \; \; \; 
  P(q_{1},q_{2}) = \bar{\psi}^{a}(q_{1}) \gamma_{5} \psi^{a}(q_{2}), 
  \; \; \; 
  V^{\mu}(q_{1},q_{2}) = \bar{\psi}^{a}(q_{1}) \gamma^{\mu} \psi^{a}(q_{2}).
\eeq  
The ERG equation is solved for the truncated action which is represented by a 
finite sum $S = S^{(2,1)} + S^{(4,0)} + S^{(4,2)} + \ldots + S^{(n,m)}$ 
and which contains all possible operators with up to $n$ fermionic fields 
and up to $m$ derivatives. Such expression can be viewed as a certain 
order of the derivative expansion with each term being further 
approximated by a polynomial in powers of fields.   
The general form of the first few $S^{(n,m)}$ respecting the symmetries 
of the theory can be shown to be
\bea
S^{(2,1)} & = & - \int \frac{d^{2}q}{(2\pi)^{2}} \bar{\psi}(-q) i\hat{q}
 \psi (q), \label{action-ferm1} \\
S^{(4,2)} & = & \int \frac{d^{2}q_{1} \ldots d^{2}q_{4}}{(2\pi)^{8}} 
 (2\pi)^2 \delta (\sum q_{i}) \left[ g_{1}(t) (S(q_{1},q_{2})S(q_{3},q_{4}) - 
 P(q_{1},q_{2})P(q_{3},q_{4}) )  \right. \nonumber \\
   & + & \left. g_{2}(t) V^{\mu}(q_{1},q_{2}) 
 V^{\mu}(q_{3},q_{4}) \right],  \nonumber \\
 S^{(4,2)} & = & \int \frac{d^{2}q_{1} \ldots d^{2}q_{4}}{(2\pi)^8} (2\pi)^2
 \delta (\sum q_{i}) \left[ \left(m_{1}(t) (q_{1}+q_{2})^2 + m_{2}(t) 
 (q_{1}+q_{2}) (q_{3}-q_{4}) \right. \right.  \nonumber \\ 
  & + & \left. \left. m_{3}(t) (q_{1}-q_{2})^2 \right) 
 \left( S(q_{1},q_{2})S(q_{3},q_{4}) - P(q_{1},q_{2})P(q_{3},q_{4}) \right)
 + \ldots \right].         \nonumber
\eea
The coefficients $g_{i}(t)$, $m_{i}(t)$, etc. are running coupling 
constants of various operators. Within such approximation the ERG equation 
(\ref{ERG-ferm}) becomes a system of equations for these coupling 
constants.  
$S^{(2,1)}$ is of course the first term of the expansion of the 
kinetic part of (\ref{action-ferm}) in powers of momenta (as before, we use 
normalization $K(0)=1$). We assume that the normalization of the kinetic 
term is fixed to be the canonical one, so no $t$-dependent coefficient appears 
here.  

The action with just $S^{(2,1)}$ and $S^{(4,0)}$ terms of 
(\ref{action-ferm1}) 
is the chiral Gross-Neveau model \cite{GN}. This model was studied in a 
number of papers and was shown to have non-trivial fixed points which 
cannot be obtained neither by perturbative methods nor in the $1/N$ 
expansion \cite{DF}. One fixed point corresponds to 
the abelian Thirring model with $g_{1}=0$ and $g_{2}$ arbitrary. 
For the second 
one the coupling $g_{1} = 4\pi /(N+1)$ and $g_{2}$ is again arbitrary. 

It can be shown that the lowest order approximation which gives a  
fixed point solution with non-zero value for the anomalous dimension must  
contain at least 6 fermions and 3 derivatives. Precisely 
such action $S = S^{(2,1)} + S^{(2,3)} + S^{(4,0)} + S^{(4,2)} + 
S^{(6,1)} + S^{(6,3)}$
was used for calculations in Ref. \cite{CKM}. It involves 107 operators 
in total: 1 with 2 fermions and 
1 derivative (the kinetic term), 1 in $S^{(2,3)}$, 2 in $S^{(4,0)}$, 
11 in $S^{(4,2)}$, 5 in $S^{(6,1)}$ and 87 in $S^{(6,3)}$. It turns 
out that the coupling constant of the operator with 2 fermions and 
3 derivatives decouples 
from the rest of the system and does not play any role.  
With this action fixed point solutions were found and corresponding 
critical exponents were calculated. 
The results are as follows. 

First of all, consistent limits $N \rightarrow \infty$ of the system of 
the fixed point equations can be considered. In this limit the 
system simplifies dramatically. Two different 
large-$N$ regimes of the behaviour of the coupling constants were found and 
correspondingly one fixed point solution for each regime was obtained 
(let us call them fixed points I and II). 

Second, for finite $N > 1$  among numerous solutions (most of which 
are spurious) two sequences of fixed points in the space of  
coupling constants were identified. Namely, as $N$ grows these solutions 
match fixed point I and II. For each fixed point, both 
for finite $N$ and for $N=\infty$, the corresponding anomalous 
dimension $\eta$ and the critical exponent $\lambda_{1}$ 
of the most relevant operator were calculated. The fixed point solutions 
are represented by their values of $\eta$ and $\lambda$ in Fig. 1 and Fig. 2.

\vskip 0.1cm 
\begin{figure}[ht]
\epsfxsize=0.9\hsize
\epsfbox{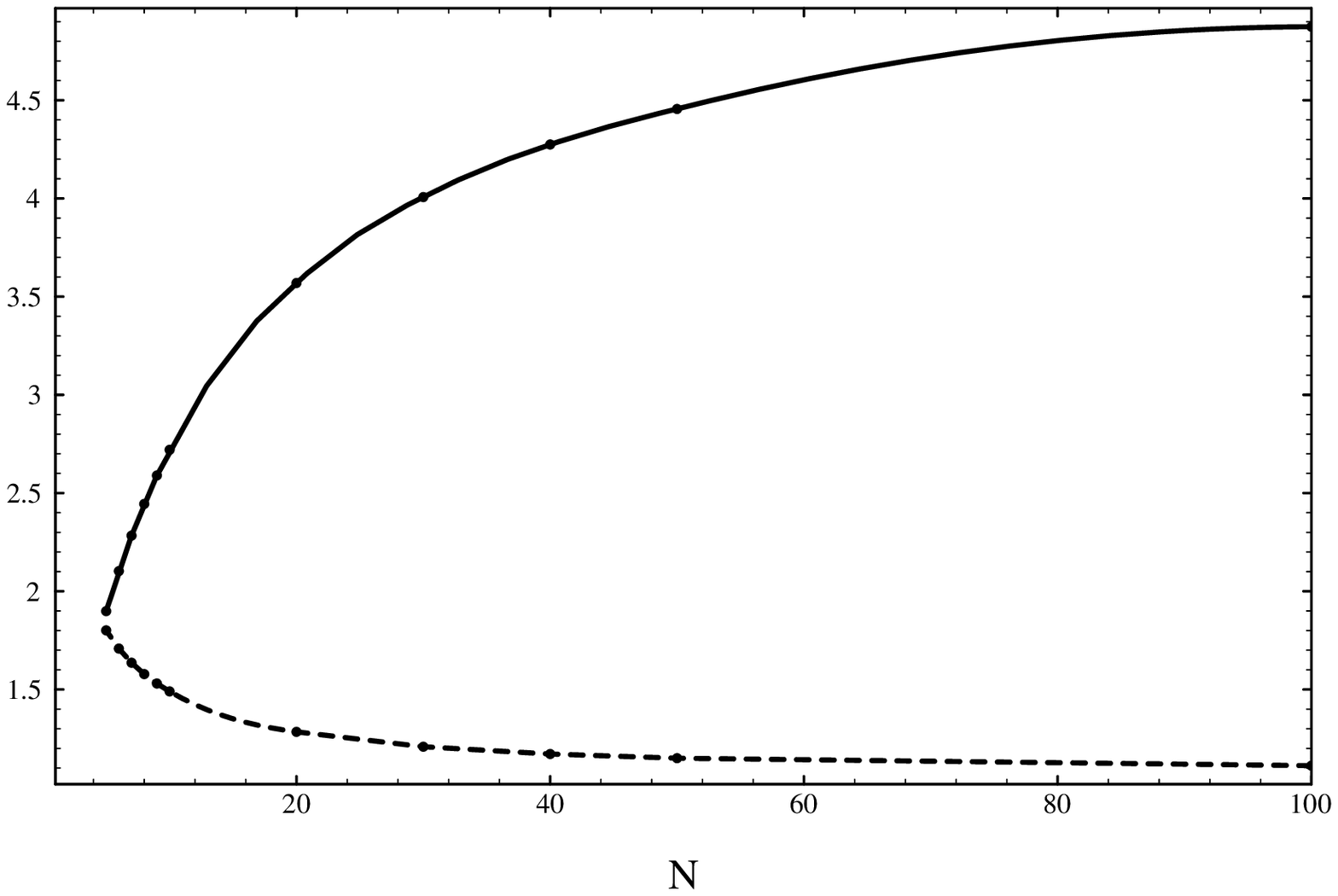}
\caption{The values of $ N \eta$ (solid line) and $\lambda_{1}$ (dashed line) 
for the sequence of fixed points approaching type I fixed point of the 
$N=\infty$ case.}
\end{figure}

\vskip 0.1cm 
\begin{figure}[ht]
\epsfxsize=0.9\hsize
\epsfbox{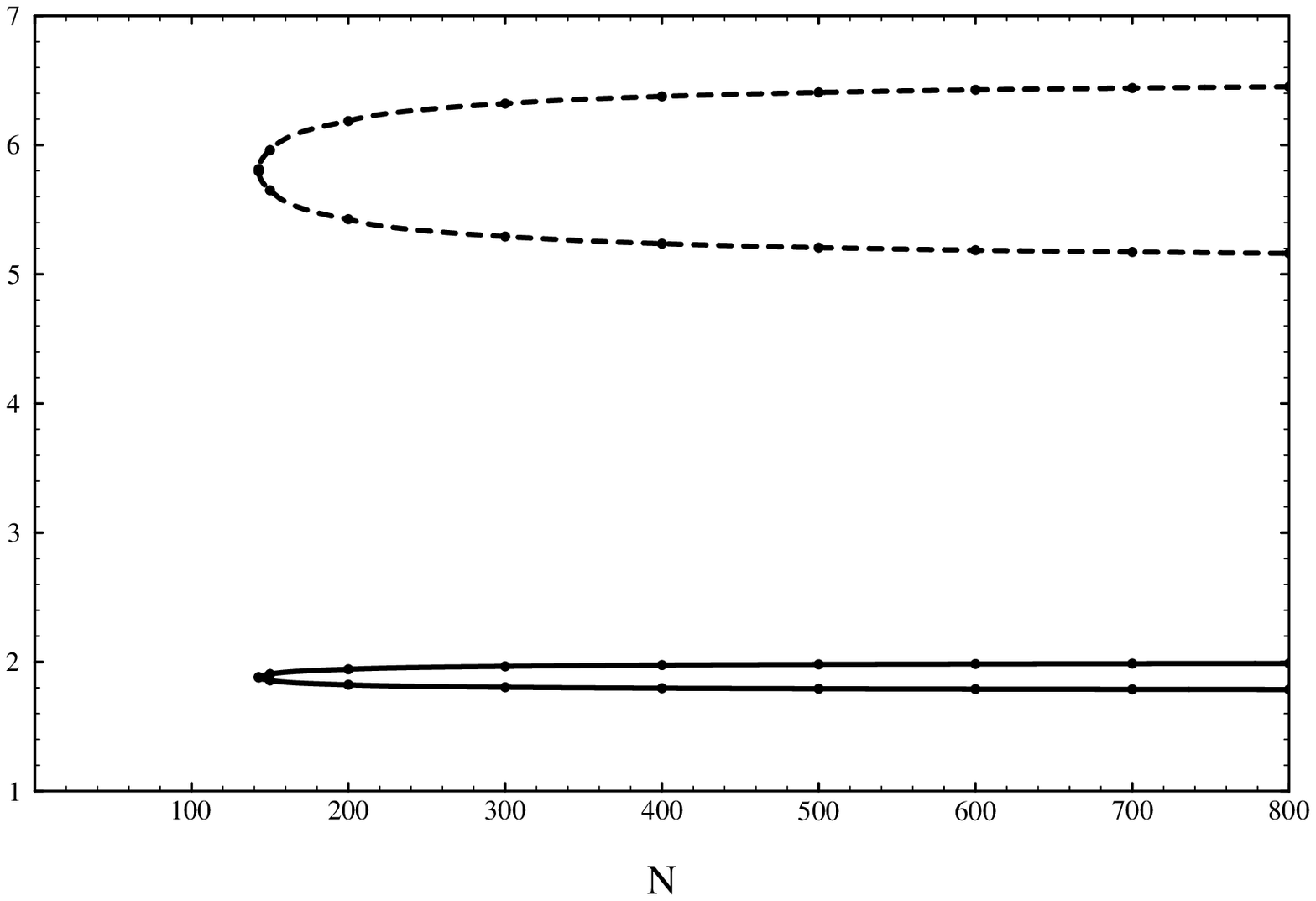}
\caption{The values of $\eta$ (solid line) and $\lambda_{1}$ (dashed line) 
for the second sequence of fixed points. The upper branches approach the 
corresponding values of type II fixed point of the 
$N=\infty$ case.}
\end{figure}

For the sequence of solutions, which approach fixed point I, $\eta 
\sim 1/N$, whereas $\lambda_{1} $ approaches the finite limit $\lambda_{1}=
1.231...$  (Fig. 1). The value of $\lambda_{1}$ for $N=\infty$ is 
scheme independent 
for a wide range of scheme parameters. For this fixed point the constants 
$g_{2}$ and $m_{1}$ remain unfixed, a feature which resembles the 
exact result by Dashen and Frishman \cite{DF}. For finite $N$ the value of 
$\lambda_{1}$ depends on the scheme, so the calculation was done for the 
regulating function $K(z)=\exp (-z^2)$.  

The second sequence of solutions, the one which matches fixed point II, 
corresponds to the upper branches in Fig. 2. It exists only for $N \geq 143$. 
At the value $N=142.8$ it joins another 
line of fixed point solutions. The values 
of $\eta$ and $\lambda_{1}$ are scheme dependent and were fixed in some 
cases by the minimal sensitivity criterion. 
All coupling constants are fixed for these fixed points. 

The case $N=1$ was considered separately. In this case due to the Fierz 
relations the number of independent operators reduces. For example,  
all operators in $S^{(6,1)}$ vanish because  
of the Grasmannian nature of the fermionic fields. In this case the 
approximation under consideration becomes the {\it complete} 
derivative expansion 
with terms up to 3 derivatives, i.e. there are no other operators with 1, 2 
and 3 derivatives apart from the ones described above. In this case the 
system of the ERG equations has just one fixed point solution. Values of 
all constants are fixed for this solution. 

Let us note that in this respect the case of finite $N > 1$ is qualitatively 
different: in this case the approximation considered here is a 
truncation of the  
derivative expansion in the number of fields. In this case one observes 
a big number 
of fixed point solutions for each $N$. Most of them are, of course, 
spurious, similar to the polynomial approximation in the scalar case. 
Since successive approximations were not studied, it was not possible 
to apply the criterion of stability of solutions for various approximations, 
as it was done in the scalar theory \cite{HKLM} . So, only those fixed 
points which form a sequence of solutions matching type I or type II 
solutions when $N \rightarrow \infty$ were chosen. 

\section{Conclusions}

We would like to finish with a few conclusive remarks.

1. The RG functional equations of type (\ref{func}) 
and the ERG equation (\ref{ERG-eq}) both reflect functional self-similarity 
of corresponding quantities and, thus, have very much in common 
at the fundamental level. The derivation 
of the RG equation essentially relies on the multiplicative character 
of finite Dyson transformations.  They appear because of the freedom in 
fixing of finite arbitrariness which is left after the removal of 
ultraviolet divergencies. This means, first, that some underlying perturbative 
expansion in a coupling constant (or analyticity in the coupling 
constant) is assumed and, second, that by construction the RG approach  
is formulated for renormalizable theories with the upper cutoff $\Lambda_{0}$  
being sent to infinity and, thus, all irrelevant operators being removed. 
This is not the case for ERG approach where the cutoff is kept finite. 
Thus, the built-in functional self-similarity of functional RG 
equations of type (\ref{func}) imposes strong 
restrictions and reduce the number of arguments of $\bar{g}(x,g)$, 
but these equations do not contain enough of the dynamical 
information to determine 
the function \cite{BSh}. The only regular way to proceed is to use 
renormalization group functions (Gell-Mann-Low $\beta$-function, etc.) 
calculated within the perturbation theory where contributions of 
irrelevant operators enter through loop corrections. 
Contrary to this, the ERG 
equation (\ref{ERG-eq}) basically contains information about renormalization 
group evolution of the whole Wilson effective action, i.e. about 
all operators and, as such, does not need any 
additional inputs. This allows to search for non-perturbative 
(i.e. non-analytical in the coupling constant) solutions 
\footnote{I thank J.I. Latorre for attracting my attention 
to these issues and clarifying some of them.}.    
    
\noindent 2. The ERG method was demonstrated to be a powerful approach for 
non-perturbative studies of the continuum limit in quantum field 
theory for scalar and fermionic theories.

\noindent 3. The proper derivative expansion is an 
effective and reliable technique 
which allows to search for fixed point solutions and calculate 
critical exponents. When combined with further polynomial 
approximation, some qualitative features can be reproduced. However, 
numerous spurious solutions appear and a special procedure should be 
applied to identify the true fixed points.

\noindent 4. For ERG equations with an arbitrary cutoff function $K(z)$ 
results beyond the leading ${\cal O}(\partial^{0})$ approximation 
depend on $K(z)$. This regulator dependence is similar to the 
scheme dependence in usual perturbation theory in the RG approach.
This is also related to an important issue of reparametrization 
invariance. We did not consider it here, and the readers are referred to 
other articles where this problem is discussed \cite{repinv,Mo8}.

\noindent 5. In spite of some progress in the ERG approach in gauge 
theories \cite{gauge1}, further developments should be made before 
we have a regular tool for obtaining non-perturbative quantitative 
results for this class of theories. 

\section*{Acknowledgements}

I thank Jose Gaite, Tim Morris and Chris Stephens 
for many interesting and illuminating discussions. I am grateful to 
Jordi Comellas, Jos\'e Ignacio Latorre and Dmitri Vasilievich Shirkov 
for reading the paper and 
for their valuable and helpful remarks. The work was supported by 
the grants INTAS-93-1630.EXT and CERN/P/FAE/1030/95 (JNICT, Portugal) 
and by CIRIT (Generalitat de Catalunya).

\end{document}